\documentclass[preprint,showpacs,preprintnumbers,amsmath,amssymb,prb,aps]{revtex4-1}
\usepackage{epsfig}
\usepackage{graphicx}
\usepackage{dcolumn}
\usepackage{bm}
\usepackage{textcomp}

\begin{document}

\title{Limitations of unconstrained LSDA+$U$ calculations in predicting the electronic and magnetic ground state of a geometrically frustrated ZnV$_{2}$O$_{4}$ compound}
\author{Sohan Lal} 
\altaffiliation{Electronic mail:goluthakur2007@gmail.com}
\author{Sudhir K. Pandey}
\altaffiliation{Electronic mail:sudhir@iitmandi.ac.in}
\affiliation{School of Engineering, Indian Institute of Technology Mandi, Kamand 175005, Himachal Pradesh, India}

\date{\today}

\begin{abstract}

   In the present work, we investigate the applicability of the LSDA+$U$ method in understanding the electronic and magnetic properties of a geometrically frustrated ZnV$_2$O$_4$ compound, where the delicate balance of electrons, lattice, orbital and spin interactions play an important role in deciding its physical properties. In the ferromagnetic solution of the compound, only one type of orbital solution is found to exist in all ranges of $U$ studied here. However, in antiferromagnetic (AFM) phase, two types of orbital solutions, AFM(OS1) and AFM(OS2), exist for $U >$3 eV. If the difference of the electronic occupancy of $d_{xz}$ and $d_{yz}$ orbitals is less than 0.25, then AFM(OS1) solution is stabilized, whereas for higher values AFM(OS2) solution is stabilized. The use of unconstrained calculations within the fully localized double counting scheme is unable to predict the AFM ground state for $U \leqslant$3 eV. Our results clearly suggest the importance of constrained calculations in understanding the electronic and magnetic properties of a compound, where various competing interactions are present. In the AFM solution, the orbital ground state of the compound changes with varying $U$, where AFM(OS1) is found to be the ground state for $U \leqslant$3 eV and for higher values of $U$, AFM(OS2) is the ground state. The analysis of the band gap suggests that the AFM(OS2) is the real ground state of the compound.  
              
\end{abstract}


\maketitle

\section{Introduction} 
   
    Theoretically, it has always been a challenging task to know the true ground state of a correlated electron system. The task becomes more daunting, when the system possesses various competing interactions. The density functional theory (DFT) has been remain one of the prominent theoretical tools to study the ground state properties of various materials. \cite{Lundqvist,Mahan,Martin,Singh} The ground state properties of the metallic and semiconducting systems are normally described by considering local spin density approximation (LSDA) or generalized gradient approximation (GGA) exchange correlation functional.\cite{Wang,Burke,Perdew} The ground state properties of strongly correlated systems, are described by adding orbital dependent Hartree-Fock (HF) potential to the LSDA/GGA functional, which is known as a LSDA+$U$/GGA+$U$ approximation.\cite{Anisimov,Liechtenstein,Solovyev,Bultmark} In this approximation, one has to deal with the double counting (DC) of the interactions. In this situation the best way is to identify and subtract the mean-field-part of the HF potential. Czy$\dot{\rm z}$yk and Sawatzky have proposed a scheme that is applicable for the weakly correlated system, which is known as around mean field (AMF) DC scheme.\cite{Bultmark,Czyzyk} For strongly correlated systems, AMF DC scheme may not be valid. In these systems, one can prefer the fully localized (FL) DC scheme, where the average effect for localized states is subtracted with integer occupation numbers.\cite{Liechtenstein,Bultmark} It is also well known that LSDA+$U$ calculations often converge to the local minima and hence they may predict the wrong ground state of the system.\cite{Korotin,Knizek,Dorado,Pandey2010} In order to know the magnetic ground state of a system, one has to compare the energy of various spin configurations. If no constraint is applied to the moments of the magnetic ions, it has normally been seen that the magnetic moment (MM) inside the muffin-tin sphere of ferromagnetic (FM) and antiferromagnetic (AFM) solutions is slightly different. In such a situation, a small change in MM may predict the wrong ground state of the system, where several kinds of competing interactions among various degrees of freedom are present. 
    
    From above discussions, it is clear that the normal LSDA+$U$ calculations are prone to predict the wrong ground state of the complex systems, where various competing interactions are present. Spinel vanadates, having general formula AV$_{2}$O$_{4}$ (A=Zn, Cd and Mg) are such complex systems, which have attracted a great deal of attention for last 15 years.\cite{Lee,Mamiya,Nishiguchi,Onoda,Reehuis,Tsunetsugu,Tchernyshyov,Motome,Matteo,Radaelli,Maitra,Wheeler,Pandey2011,Pandey2012,Lal} At room temperature magnetic V ions of these compounds form a geometrically frustrated pyrochlore lattice due to corner sharing tetrahedral network.\cite{Lee,Suzuki} The magnetically active V$^{3+}$ ions have two 3$d$ electrons in $t_{2g}$ orbitals forming $S$=1 state, which are antiferromagnetically coupled to each other. These compounds undergo the structural and magnetic transitions from cubic to tetragonal and paramagnetic to antiferromagnetic, respectively, at two different temperatures and show qualitatively similar structural and magnetic behavior.\cite{Mamiya,Nishiguchi,Onoda,Reehuis,Radaelli} The structural and magnetic transitions observed in these compounds were intriguing for a long time, as compounds with geometrically frustrated pyrochlore lattice normally do not show any structural and magnetic transitions. Because of the presence of 2 electrons in the $t_{2g}$ orbitals, V ions become Jahn-Teller active and hence one expects its effect on the structure. Many groups proposed that the orbital ordering is the key factor responsible for low temperature structural transitions observed in these compounds.\cite{Tsunetsugu,Motome,Tchernyshyov,Radaelli,Huang} Due to this structural transition, geometrical frustration gets relaxed and hence these compounds show magnetic transitions. However, the non-coincidence of the structural and magnetic transitions also suggest the presence of a certain degree of geometrical frustration in these compounds. The experimentally observed values of the magnetic moments are found to be much smaller than the expected value of 2 $\mu_B$, which has been attributed either to the presence of the large but negative value of orbital momentum or to the presence of a certain degree of geometrical frustration.\cite{Reehuis,Maitra,Wheeler,Pandey2011,Pandey2012} These discussions clearly show that the structural, electronic and magnetic properties of these compounds are decided by the complex interplay of the electron, lattice, spin and orbital degrees of freedom. 
    
    Most of the theoretical studies carried out on these compounds are based on the model calculations, which are fully parameter dependent. There are very few LSDA+$U$ based first principles studies available in the literature, where only two free parameters $U$ and $J$ are used. In spite of the wide applicability of this method in studying the ground state properties of correlated electron system, there are no systematic studies available in the literature, which show the applicability of the LSDA+$U$ method in verifying the experimentally observed ground state of these complex oxides. In this work, we have made such an attempt on ZnV$_2$O$_4$, which is the most studied compounds among the spinel vanadates. The experimentally observed magnetic structure of ZnV$_{2}$O$_{4}$ is AFM. The crystal structure of the compound consists of edge sharing octahedra both along the $a$ and $b$-axes. Along these directions, spins are antiferromagnetically aligned in sequence $\uparrow$$\downarrow$$\uparrow$$\downarrow$. However, the antiferromagnatically coupled chains along the $a$ and $b$-axes are connected to each other by two FM and two AFM bonds.\cite{Reehuis,Niziol,Radaelli} ZnV$_2$O$_4$ compound has an insulating ground state with a band gap $\sim$ 0.32 eV.\cite{Rogers} It shows structural and magnetic transitions at $\sim$ 50 and 40 K, respectively.\cite{Lee} The experimentally observed MM per V atom is found to be $\sim$ 0.63 $\mu_B$.\cite{Reehuis} This is reported to be the most frustrated compound among the spinel vanadates with the frustration index of $\sim$ 21.\cite{Pandey2012}   
 
    Here, we report the applicability of the LSDA+$U$ method in understanding the electronic and magnetic properties of ZnV$_2$O$_4$ compound. This work clearly suggests the inadequacy of the unconstrained LSDA+$U$ calculations in predicting the AFM ground state for a wide range of $U$. The constrained LSDA+$U$ calculations provide correct AFM ground state of the compound in all ranges of $U$ studied here. The FM solution gives only one type of orbital solution (OS1). In the AFM phase, two types of orbital solutions, AFM(OS1) and AFM(OS2) exist for $U$ $>$3 eV and only AFM(OS1) is found to exist for $U \leq$3 eV. The ground state of the compound with varying $U$ depends on the values of the electronic occupancy difference of $d_{xz}$ and $d_{yz}$ orbitals ($\mid$$d_{xz}$-$d_{yz}$$\mid$). For $\mid$$d_{xz}$-$d_{yz}$$\mid$ $<$0.25, AFM(OS1) has the lowest energy and for higher values of $\mid$$d_{xz}$-$d_{yz}$$\mid$, AFM(OS2) becomes the ground state of the system. The comparison of the experimental band gap with the calculated one suggests that the AFM(OS2) is the true ground state of the compound.

\section{Computational Detail}       
       
    The FM and AFM calculations of ZnV$_{2}$O$_{4}$ have been carried out by using the {\it state-of-the-art} full-potential linearized augmented plane wave (FP-LAPW) method, as implemented in elk code.\cite{elk} All calculations are carried out by using unconstrained and constrained LSDA+$U$ method.\cite{Bultmark} In this method, one has to deal with the DC of the interactions, as the electron-electron interactions have already been included in LSDA potential. In this work, we have used the FL and AMF DC schemes, which are applicable to strongly and weakly correlated systems, respectively.\cite{Czyzyk,Liechtenstein} The corrected total energy functional for FL DC scheme is written as:\cite{Liechtenstein}
   
 \begin{eqnarray}
E^{{\rm LSDA}+U}[\rho^{\sigma}({\bf r}),\{n^{\sigma}\}]&=&E^{{\rm LSDA}}[\rho^{\sigma}({\bf r})]+\frac{1}{2} \sum_{\{m\},\sigma}\{\langle m,m''|V_{ee}|m,m''\rangle n_{mm'}^\sigma n_{m''m'''}^{-\sigma} \nonumber \\
&&+(\langle m,m''|V_{ee}|m',m'''\rangle-\langle m,m''|V_{ee}|m''',m'\rangle)n_{mm'}^\sigma n_{m''m'''}^{\sigma}\} \nonumber \\
&&-\frac{1}{2}Un(n-1)+\frac{1}{2}J[n^\uparrow(n^\uparrow-1)+n^\downarrow(n^\downarrow-1)]     
\end{eqnarray}

    where, $E^{\rm LSDA}$[$\rho$$^{\sigma}$({\bf r})], $\rho$$^{\sigma}$({\bf r}), and $V_{ee}$ are the standard LSDA functional, the charge density for spin-$\sigma$ electrons and the screened Coulomb interactions among the $nl$ ($n$ and $l$ are the principal and orbital quantum, respectively) electrons, respectively. $n_{mm'}^\sigma$ and $n_{m''m'''}^{\sigma}$ are the ($m$, $m'$ ) and ($m''$, $m'''$) elements of the density matrix with spin $\sigma$ for correlated electrons, respectively. $U$, $J$, $n^\sigma$=Tr($n_{mm'}^{\sigma}$) and $n$=$n^\uparrow$+$n^\downarrow$ are screened Coulomb parameter, exchange parameter, total number of electrons with spin $\sigma$, and total number of electrons, respectively. 
   
Similarly for AMF DC scheme, the corrected total energy functional is written as:\cite{Czyzyk}
    
\begin{eqnarray}
E^{{\rm LSDA}+U}&=&E^{{\rm LSDA}}+\frac{1}{2} \sum_{m,m',\sigma} U_{m,m'} n_{m\sigma} n_{m'-\sigma}+\frac{1}{2} \sum_{m,m',m\neq m',\sigma} (U_{m,m'}-J_{m,m'}) n_{m\sigma} n_{m'\sigma}\nonumber \\
&&-\frac{1}{2}UN(N-1)+\frac{1}{2}J[N^\uparrow(N^\uparrow-1)+N^\downarrow(N^\downarrow-1)]     
 \end{eqnarray}   
   
   where, $E^{\rm LSDA}$ and $N$=$N^\uparrow$+$N^\downarrow$ are the standard LSDA functional and the total number of electrons, respectively. ($n_{m\sigma}$, $n_{m'\sigma}$) and ($U_{m,m'}$, $J_{m,m'}$) are the charge density of the orbitals and matrices of the screened Coulomb parameter $U$ and exchange parameter $J$, respectively. 
       
    ZnV$_{2}$O$_{4}$ crystallizes in the body centered tetragonal spinel structure with the space group $I4_{1}/amd$. The experimentally observed AFM structure, the lattice parameters and atomic positions are taken from the literature.\cite{Reehuis} In order to check the robustness of the method, it is desirable to do the full structure optimization using both unconstrained and constrained LSDA+$U$ calculations in understanding the electronic and magnetic properties of the compound. Such calculations are very much time consuming and are beyond the scope of the present work. Perdew -Wang/Ceperley -Alder exchange correlation functional is used in the calculations.\cite{Perdew} The effect of on-site Coulomb interaction between V 3$d$ electrons is considered within the LSDA+$U$ formulation of the DFT.\cite{Bultmark} Normally in LSDA+$U$ method, $U$ and $J$ are used as parameters. In this work, only $U$ is used as a free parameter and the value of $J$ is calculated self-consistently, as described in reference [11]. In order to understand the electronic and magnetic properties of the compound, we have performed collinear unconstrained and constrained magnetic calculations by varying $U$ from 0-6 eV, where the direction of MM is fixed along the z-axis. In unconstrained calculations, the magnitude of MM of every V atoms (inside the muffin-tin sphere) is allowed to vary self-consistently for every kind of solutions. However, in the constrained calculations, the value of MM of every V atoms in FM solution is kept same to the self-consistently obtained value of MM in AFM solution. The  muffin-tin sphere radii used in the calculations are 2.0, 2.0 and 1.54 Bohr for Zn, V and O, respectively. (6,6,6) k-point mesh size is used. The basis set cut-off of muffin-tin radius times maximum $\lvert$G+k$\lvert$ (rgkmax) and maximum length of $\lvert$G$\lvert$ for expanding the interstitial density and potential (gmaxvr) are set to be 7.0 and 12.0, respectively, in all the calculations. Convergence target of total energy has been set below ~10$^{-4}$ Hartrees/cell.
   
\section{Result and Discussion} 

   First of all, we study the applicability of unconstrained LSDA+$U$ calculations, in predicting the experimentally observed ground state of ZnV$_{2}$O$_{4}$ compound for a wide parameter range. The experimentally observed antiferromagnetic structure of the compound as determined by the Reehuis $et$ $al$. used in the calculations is shown in the Fig. 1. It is evident from the figure that each V atom is surrounded by six O atoms forming an octahedron (O-O bond is not connected for the sake of clarity). The crystal structure of the compound consists of edge sharing octahedra both along the $x$ and $y$-axes. Along these directions, spins are antiferromagnetically aligned in sequence $\uparrow$$\downarrow$$\uparrow$$\downarrow$ forming the antiferromagnatically coupled chains.\cite{Reehuis,Niziol,Radaelli} Here, it is important to note that the LSDA+$U$ calculations often converge to local minima depending upon the starting electron densities and potentials. In the present study, the FM and AFM solutions corresponding to both DC schemes were obtained by considering various combinations for electron densities and potentials, for $U$=0-6 eV, where $U$=0 provides the simple LSDA solution. For both the DC schemes and for all values of $U$, only one type of orbital solution (OS) exists in the FM solution of the compound, where every calculation corresponding to FM structure converges to only one type of orbital solution. For $U \leq$3 eV, every calculation corresponding to AFM structure converges to only one OS and is similar to that observed in FM solution. However, for $U >$3 eV, two types of OS exist, where the AFM structure for every calculation either converges to one type of orbital solution, denoted by OS1 or another type of orbital solution, denoted by OS2. Here, OS1 is similar to that obtained in the FM solution. From here onward, AFM structure with two orbital solutions, OS1 and OS2 are denoted by AFM(OS1) and AFM(OS2), respectively. The nature of both the orbital solutions is explained later in the manuscript. The structural unit cell of ZnV$_{2}$O$_{4}$ contains four Zn, eight V and sixteen O atoms, which are located at the 4$a$ (0, 3/4, 1/8), 8$d$ (0, 0, 1/2) and 16$h$ (0, 0.0200, 0.2611) Wyckoff position, respectively. All the eight V atoms are structurally equivalent because of the same Wyckoff position with the underlying space group symmetry of the crystal ($I4_{1}/amd$). However, they are not orbitally equivalent. The structural unit cell along with the spin arrangements of eight V atoms is shown in Fig. 2. According to the orbital occupancy, these V atoms can be divided into two groups containing (V1,V2,V5,V6) and (V3,V4,V7,V8) atoms, and mainly occupying the $d_{xz}$ and $d_{yz}$ orbitals, respectively in the global coordinate system. Hence, such type of preferred occupation of these orbitals leads to the anti-ferro orbital ordering (OO) for every kind of solutions. OO of this compound is studied by different groups. Here, we compare our result with the results of other groups. Based on the three-dimensional electron density plot of ZnV$_2$O$_4$ compound calculated within LSDA+$U$, Maitra $et$ $al$. have shown the $d_{xz}$+$d_{yz}$ and $d_{xz}$-$d_{yz}$ OO.\cite{Maitra} However, based on model calculations, Tsunetsugu $et$ $al$. have predicted $d_{xz}$ and $d_{yz}$ OO.\cite{Tsunetsugu} OO obtained by these groups is in the local octahedral coordinate system, where they have set x and y axes along the two basal V-O bonds and z axis along the apical V-O bond of regular VO6 octahedron. However, in the present study, we have used the global coordinate system, where x and y axes are making angles of 47.4 degree and 42.6 degree with two basal V-O bonds and z axis is making an angle of 4.8 degree with the apical V-O bond. This is because of the distorted VO6 octahedron in the tetragonal phase of the compound. If we rotate these axes (local octahedral coordinate system), the OO in the present study will be similar to that obtained by Maitra $et$ $al$.\cite{Maitra} The OO predicted by the Tsunetsugu $et$ $al$. is inconsistent with spatial symmetry of $I4_1/amd$ space group, as predicted by x-ray scattering experiments on polycrystalline samples.\cite{Reehuis,Nishiguchi} Two orbital solutions (with $d_{xz}$ and $d_{yz}$ OO in the global coordinate system) obtained in the present study are compatible with the crystal symmetry as both solutions are obtained by the self-consistent field calculations on ZnV$_{2}$O$_{4}$ compound described by the space group $I4_1/amd$. For more details, the readers are advised to see the reference [31], where we have shown the OO in both coordinate systems for this compounds. In order to know the magnetic ground state of the compound, we have compared the total energy of FM and AFM solutions, for all values of $U$ studied here. The energy difference ${\Delta}E$=$E_{AFM}$-$E_{FM}$ corresponds to both the DC schemes with varying $U$ is plotted in Fig. 3, where ${\Delta}E_1$=$E_{AFM(OS1)}$-$E_{FM}$ and ${\Delta}E_2$=$E_{AFM(OS2)}$-$E_{FM}$.
  
   It is evident from Fig. 3(a) that the ground state predicted by both the DC schemes is not consistent for $U$=1-3 eV, as FL and AMF DC schemes give FM (not experimental one) and AFM(OS1) ground state, respectively. Above $U$=3 eV, both the DC schemes show FM ground state, which is not the experimentally observed ground state of the compound. Interestingly, both the DC schemes provide AFM(OS2) ground state for $U >$3 eV, as evident from Fig. 3(b). Thus, AFM(OS2) appears to be the true ground state of the compound for $U >$3 eV. However, for $U$=1-3 eV, FL DC scheme fails to provide correct magnetic ground state. Keeping the earlier report about the strength of on-site Coulomb interaction\cite{Canosa} in mind, one may come to the conclusion that the FL DC scheme is not suitable for predicting the correct ground state of the compound. However, the careful analysis of the data suggests that the cause of this discrepancy lies in the unconstrained nature of the calculations, which is clear from the following discussion. The MM per V atom of FM (M$_{\rm FM}$) and  AFM(OS1) (M$_{\rm AFM(OS1)}$) solutions for the FL DC scheme is shown in Table I, for $U$=1-6 eV. The values of M$_{\rm FM}$ and M$_{\rm AFM(OS1)}$ are different because of the unconstrained nature of the calculations. The difference between these two moments $\Delta$M, decreases from 0.178 to 0.06 $\mu$$_{B}$, when $U$ is increased from 1 to 6 eV. The value of $\Delta$M is $>$0.13 $\mu$$_{B}$, for $U \leq$3 eV. At this stage, one may conjecture that the observed extra MM per V atom in FM solution as compared to AFM solution appears to responsible for making the total energy of the FM solution less than that of AFM for $U$=1-3 eV. At this point, it is important to note that AMF DC scheme also provides similar values of $\Delta$M with varying $U$. However, in spite of this discrepancy, the AMF DC scheme seems to predict the correct AFM ground state of the compound, as the energies of AFM solutions are less than the FM solutions, where AFM structure is the experimentally observed ground state of the compound.  
   
  In order to verify the above conjecture, we have performed the constrained calculations, where the magnitude of the MM of each V atoms is kept same for every type of solutions as described in computation details section. The comparison of energy obtained from these calculations is shown in Fig. 4, where ${\Delta}E_1$=$E_{AFM(OS1)}$-$E_{FM}$ and ${\Delta}E_2$=$E_{AFM(OS2)}$-$E_{FM}$. It is evident from Fig. 4(a) that the AMF DC scheme provides correct AFM ground state for all values of $U$ studied here, whereas FL DC scheme predicts AFM(OS1) ground state for $U \leqslant$4 eV. Above $U$=4 eV, FL DC scheme gives FM ground state, which is not as per experimental result. Interestingly, both the DC schemes predict AFM(OS2) as a ground state, which is evident from Fig. 4(b). These results show the drastic improvement in predicting the AFM ground state of the compound, when constraint on the magnetic moments of V atoms is invoked. 
    
   As mentioned above, for $U >$3 eV, two orbital solutions OS1 and OS2 exist in the AFM phase of the compound. In order to know the true ground state of the compound for $U >$3 eV, we have calculated, ${\Delta}E$=$E_{AFM(OS2)}$-$E_{AFM(OS1)}$. The ${\Delta}E$ vs $U$ plot for both the DC schemes are shown in Fig. 5. This graph shows that the AFM(OS2) is the true ground state of the compound as we got larger, but negative values of ${\Delta}E$ for $U \geq$4 eV. Moreover, $\Delta$$E$ decreases continuously with increase in $U$ from 4-6 eV. The decrease in the $\Delta$$E$ with increasing $U$ gives rise to more stability to the OS2 as compared to OS1. Thus, the present work shows the capability of both DC schemes, in predicting the experimentally observed AFM ground state on invoking the constraint calculations. This work also suggests that for $U \leq$3 eV, AFM(OS1) is the ground state due to the existence of only one type of OS1 solution in the AFM phase, whose energy is less than the FM solution. However, for $U >$3 eV, AFM(OS2) is the true ground state of the system, as its energy is less than both AFM(OS1) and FM solutions. From the above discussion, it is clear that the limitations of unconstrained calculations in predicting the true ground state of the complex systems can be overcome by performing the constrained calculations.             
    
   In order to know the nature of two orbital solutions OS1 and OS2 found in the AFM phase of the compound, we have plotted the electron occupancy of $d_{xy}$, $d_{xz}$ and $d_{yz}$ orbitals corresponding to $d{_\uparrow}$ electrons of the orbitally equivalent V atoms (i.e. V1,V2,V5,V6) in Fig. 6, for both the DC schemes. It is evident from the figure that in OS1 solution the occupancy of $d_{xz}$ orbital is larger than that of $d_{yz}$, whereas in OS2 solution the opposite behavior is observed. Both DC schemes provide almost the same value of orbital occupancy in OS1 solution. However, in OS2 solution, FL DC scheme gives higher occupancy for $d_{yz}$ orbital in comparison to AMF DC scheme. The occupancy of $d_{xz}$ orbital is almost same in both DC schemes. The occupancy of $d_{xy}$ orbital at every site of V atoms is almost same in OS1 solution for both DC schemes. Similar behavior is also observed in OS2. However, the occupancy of $d_{xy}$ orbital in OS1 solution is greater than that of the OS2 solution for both DC schemes. For OS1, the occupancy of $d_{xy}$ orbital is almost equal and less than $d_{xz}$ orbital for FL and AMF DC schemes, respectively. For OS2, the occupancy of $d_{yz}$ orbital is always larger than $d_{xy}$ orbital. Higher occupancy of $d_{xz}$ or $d_{yz}$ orbital as compared to $d_{xy}$ orbital is expected, as it depends on the strength of the splitting of $d_{xz}$ and $d_{yz}$ orbitals. However, the exact cause of the splitting of $d_{xz}$ and $d_{yz}$ orbitals is debated, as various groups have proposed different mechanisms.\cite{Tsunetsugu,Tchernyshyov,Matteo,Khomskii} The absolute value of electron occupancy difference between $d_{xz}$ and $d_{yz}$ orbitals ($\mid$$d_{xz}$-$d_{yz}$$\mid$) is also plotted in Fig. 6. In the OS1 solution, the $\mid$$d_{xz}$-$d_{yz}$$\mid$ shows non-monotonic behavior for both the DC schemes, where it first increases from $\sim$ 0.19 to $\sim$ 0.23 for $U$=1-3 eV and then decreases at $U$=4 eV. Above $U$=4 eV, $\mid$$d_{xz}$-$d_{yz}$$\mid$ has again started increasing. For $U \leq$3 eV (only OS1 exists), the difference in the orbital occupancy of $d_{xz}$ and $d_{yz}$ orbitals is low for both DC schemes, which is almost in close agreement with the results of Kato $et$ $al$. and Kuntscher $et$ $al$.\cite{Kato,Kuntscher} In case of OS2 solution, $\mid$$d_{xz}$-$d_{yz}$$\mid$ increases monotonically. In FL (AMF) DC scheme, it increases from $\sim$ 0.35 (0.28) to $\sim$ 0.39 (0.39) for $U$=4-6 eV. These results clearly show the discontinuous change in the value of $\mid$$d_{xz}$-$d_{yz}$$\mid$ between $U$=3 and 4 eV, where the ground state of the system changes from AFM(OS1) to AFM(OS2) with varying $U$. This point has been made more clear in Fig. 7, where we have shown $\mid$$d_{xz}$-$d_{yz}$$\mid$ with varying $U$. This figure suggests the stabilization of AFM(OS1) ground state for $\mid$$d_{xz}$-$d_{yz}$$\mid$ $<$0.25 and for higher values, AFM(OS2) ground state is stabilized.                

    The existence of two orbital solutions in AFM phase for higher $U$ may not be surprising, as it is well known that the magnetic structure of strongly correlated electron system is decided by the orientation of the orbitals in the space. The Goodenough-Kanamori-Anderson rules deal with this particular aspect of the strongly correlated electron system. This is expected to occur because of the presence of various interactions, where the delicate balance of spin, orbital, charge and lattice degrees of freedom decides the electronic and magnetic properties of the compounds. At this stage, it is difficult to identify some parameters, which are responsible for the existence of two orbital solutions. However, one may get some insight by looking at the electron occupancies of $d$, $d{_\uparrow}$, and $d{_\downarrow}$ orbitals along with the MM of the V atoms, as shown in Table II. First of all we consider OS1 solution. It is evident from the table that the total number of $d$ electrons show small $U$ dependence within FL DC scheme, as its value decreases from 2.48 to 2.42 with increasing $U$, whereas AMF DC scheme do not provide any $U$ dependence to the total number of $d$ electrons. Within FL DC scheme, the number of $d{_\uparrow}$ ($d{_\downarrow}$) increases (decreases) from  1.94 to 2.02 (0.54 to 0.40) for $U$=1 to 6 eV. Hence MM increases from 1.40 to 1.62 $\mu_B$, when $U$ increases from 1 to 6 eV. However, $d{_\uparrow}$ and $d{_\downarrow}$ electrons do not show much $U$ dependence within the AMF DC scheme and hence MM remains almost the same, which is $\sim$ 1.37 $\mu_B$. For $U \leq$3 eV, the large value of MM is observed for both DC schemes. In OS2 solution, the FL DC scheme shows a slight $U$ dependence to the total number of $d$, $d{_\uparrow}$, and $d{_\downarrow}$ electrons, which leads to the increment of MM from 1.61 to 1.66 $\mu_B$, when $U$ increases from 4 to 6 eV. The AMF DC scheme does not show any $U$ dependence to the total number of $d$ electrons. However, the number of $d{_\uparrow}$ and $d{_\downarrow}$ electrons show small $U$ dependence and leading to the increment of MM from 1.44 to 1.51 $\mu_B$, as $U$ increases from 4 to 6 eV. For higher values of $U$ (in the localized regime), the spin-orbit interaction is very effective for the large suppression of the MM, as shown by Kato $et$ $al$.\cite{Kato} The effect of the spin-orbit interaction is not considered in the present work. On comparing the MM of both the solutions, one can find that the MM of OS2 solution is always more than that of the OS1 solution. As mentioned earlier in the manuscript, such a small difference in the MM of two solutions may lead to a difference in their energy. Thus, at this stage one may attribute the negative value of ${\Delta}E$ in Fig. 5 to the higher value of MM for the OS2 solution, that may further rise the doubt about AFM(OS2), as a true ground state of the compound at higher $U$. In order to address this doubt, we have carried out the constrained calculations, where the magnitude of MM of V atoms is kept same in both the cases. The results, thus obtained show that the energy of AFM(OS2) is always less than that of AFM(OS1) for all values of $U$ and for both the DC schemes. Thus, this work clearly suggests that the AFM(OS2) is the true ground state of the compound, where anti-ferro OO corresponding to OS2 solution stabilizes the AFM order for $U >$3 eV. These results also suggest that the reason for the existence of two orbital solutions at higher $U$ appears to be due to the existence of two orbital states with respect to the AFM ordering as explained above.                    

   Finally, we discuss the dependence of the ground state insulating gap on various parameters studied here. Both the DC schemes show similar $U$ dependence gaps in AFM(OS1) ground state. At $U$=1 eV, the calculation gives metallic state. At $U$=2 eV, a small gap of $\sim$ 0.06 eV is opened up, which increases to $\sim$ 0.09 eV at $U$=3 eV. It is important to note that the experimentally observed insulating gap of the compound is $\sim$ 0.32 eV \cite{Rogers}, which is much larger than the highest gap obtained in AFM(OS1) solution. Thus AFM(OS1) does not appear to be the ground state of the compound, as the value of $U$ required to open up the experimental band gap will be higher than 3 eV. In the AFM(OS2) ground state, both the DC schemes show different $U$ dependence band gap. At $U$=4 eV, FL DC scheme shows a band gap of $\sim$ 0.36 eV, whereas AMF DC scheme fails to create any hard gap. At $U$=5 eV, the AMF DC scheme creates a hard gap of $\sim$ 0.51 eV, which increases to 0.93 eV at $U$=6 eV. The value of band gap within the FL DC scheme increases from $\sim$ 0.96 to 1.5 eV, when the value of $U$ changes from 5 to 6 eV. On comparing the calculated band gap with the experimental one, we can conclude that the AFM(OS2) is the real ground state of the compound and the value of $U$ of this compound is expected to be $\sim$ 4 eV. This value of $U$ appears to make sense in the light of the work of Canosa \textit{et al.}, where $d$ electrons of ZnV$_2$O$_4$ compound are put in the intermediate localization range.\cite{Canosa}

\section{Conclusions} 

   In summary, we have studied the applicability of the LSDA+$U$ method in understanding the electronic and magnetic properties of a complex system, where several types of competing interactions among electron, spin, orbital and lattice degrees of freedom exist, by considering ZnV$_2$O$_4$ compound as a case study. Our work clearly suggests that the unconstrained LSDA+$U$ calculations are not the correct methods for predicting the exact electronic and magnetic ground states of such systems. It is suggested that the constrained LSDA+$U$ calculations should be preferred, if one wants to predict the real magnetic ground state of an unknown complex system, as small change in the magnetic moments of magnetic atoms in various spin configurations may lead to the prediction of wrong ground state of the compound. The present study on ZnV$_2$O$_4$ compound shows the existence of only one type of orbital solution corresponding to ferromagnetic solution. However, in the antiferromagnetic phase, two kinds of orbital solutions AFM(OS1) and AFM(OS2), exist for $U >$3 eV and only AFM(OS1) is found to exist for $U \leq$3 eV. The ground state solution of the compound changes from AFM(OS1) to AFM(OS2) with varying $U$, which is related to the discontinuous change in the electronic occupancy difference of $d_{xz}$ and $d_{yz}$ orbitals ($\mid$$d_{xz}$-$d_{yz}$$\mid$). Such change in the ground state occurs around $U$=3 eV. The detailed analysis of the $U$ dependent band gaps suggests that the AFM(OS2) is the true ground state of the compound.        

\acknowledgments {S.L. is thankful to UGC, India, for financial support.}

\pagebreak

\section{Tables}

\begin{table}[ht]
\caption{The magnetic moment per vanadium atom of ferromagnetic (M$_{\rm FM}$) and antiferromagnetic OS1 (M$_{\rm AFM(OS1)}$) solutions along with the change in magnetic moment ($\Delta$M) of both solutions for fully localized double counting scheme within unconstrained LSDA+$U$ calculations for {\it U}=1-6 eV.}
\centering 
\begin{tabular}{p{3.5cm} p{5cm} p{5cm} p{2cm}}
\hline
\hline
{\it U} (eV)&{M$_{\rm FM}$($\mu$$_{B}$)}&{M$_{\rm AFM(OS1)}$($\mu$$_{B}$)}&$\Delta$M($\mu$$_{B}$)\\[0.5ex]
\hline
{1}&1.589&1.411&0.178\\
{2}&1.641&1.475&0.166\\
{3}&1.659&1.527&0.132\\
{4}&1.678&1.575&0.103\\
{5}&1.690&1.609&0.081\\
{6}&1.699&1.639&0.060\\ [1ex]
\hline
\hline
\end{tabular}
\label{table:the exp}
\end{table}

\begin{table}[ht]
\caption{Total number of $d$, {$d{_\uparrow}$} and {$d{_\downarrow}$} electrons along with the magnetic moment per V atom of orbitally equivalent V atoms (i.e. V1,V2,V5,V6) for AFM(OS1) and AFM(OS2) solutions with varying $U$ for fully localized and around mean field (in brackets) double counting schemes.}   
\begin{tabular}{p{1.5cm}p{2.0cm}p{2.0cm}p{2.0cm}p{2.5cm}p{2.0cm}p{2.0cm}p{2.0cm}p{2.0cm}p{2.5cm}}
\hline
\hline
{\it U} (eV)&\multicolumn{3}{c}{AFM(OS1)}&&\multicolumn{4}{c}{AFM(OS2)}\\
\cline{2-4}\cline{6-8}
&{$d$}&{$d{_\uparrow}$} &{$d{_\downarrow}$}&M($\mu_B$)&{$d$}&{$d{_\uparrow}$}&{$d{_\downarrow}$}&M($\mu_B$)& \\
\hline
1&2.48(2.49)&1.94(1.92)&0.54(0.57)&1.40(1.35)&-&-&-&-&\\
2&2.47(2.49)&1.97(1.92)&0.50(0.56)&1.47(1.36)&-&-&-&-&\\
3&2.46(2.49)&1.99(1.93)&0.47(0.56)&1.52(1.37)&-&-&-&-&\\
4&2.45(2.49)&2.00(1.93)&0.44(0.56)&1.56(1.37)&2.46(2.50)&2.03(1.97)&0.42(0.53)&1.61(1.44)&\\
5&2.44(2.49)&2.02(1.93)&0.42(0.56)&1.60(1.37)&2.44(2.50)&2.04(2.00)&0.40(0.50)&1.64(1.50)&\\
6&2.42(2.49)&2.02(1.93)&0.40(0.56)&1.62(1.37)&2.43(2.50)&2.04(2.00)&0.38(0.49)&1.66(1.51)&\\
\hline
\hline
\end{tabular}
\end{table}

\pagebreak

\section{Figure Captions:}

FIG. 1. Atomic and spin arrangements of V atoms along with the O atoms of antiferromagnetic ZnV$_2$O$_4$ compound in the unit cell. Each V atom is surrounded by six O atoms forming an octahedron (O-O bond is not connected for the sake of clarity) The spin of the V atoms are aligned in sequence $\uparrow$$\downarrow$$\uparrow$$\downarrow$ along the x and y-axes forming the antiferromagnetic chains.

FIG. 2. Atomic and spin arrangements of structurally equivalent eight V atoms along with the four Zn and sixteen O atoms of antiferromagnetic ZnV$_2$O$_4$ compound in the structural unit cell. The orbitally equivalent atoms (V1,V2,V5,V6) and (V3,V4,V7,V8) are mainly occupied by $d_{xz}$ and $d_{yz}$ orbitals, respectively. 

FIG. 3. The total energy difference between antiferromagnetic and ferromagnetic (${\Delta}E$=$E_{AFM}$-$E_{FM}$) solutions per formula unit corresponding to fully localized (FL) and around mean field (AMF) double counting (DC) schemes as a function of $U$ within unconstrained LSDA+$U$ calculations for both orbital solutions AFM(OS1) and AFM(OS2) denoted as (a) ${\Delta}E_1$=$E_{AFM(OS1)}$-$E_{FM}$ (b) ${\Delta}E_2$=$E_{AFM(OS2)}$-$E_{FM}$.

FIG. 4. The total energy difference between antiferromagnetic and ferromagnetic (${\Delta}E$=$E_{AFM}$-$E_{FM}$) solutions per formula unit corresponding to fully localized (FL) and around mean field (AMF) double counting (DC) schemes as a function of $U$ within constrained LSDA+$U$ calculations for both orbital solutions AFM(OS1) and AFM(OS2) denoted as (a) ${\Delta}E_1$=$E_{AFM(OS1)}$-$E_{FM}$ (b) ${\Delta}E_2$=$E_{AFM(OS2)}$-$E_{FM}$.

FIG. 5. The total energy difference between solutions AFM(OS2) and AFM(OS1) (${\Delta}E$=$E_{AFM(OS2)}$-$E_{AFM(OS1)}$) for fully localized (FL) and around mean field (AMF) double counting (DC) schemes within constrained LSDA+$U$ calculations for $U \geq$4 eV.

FIG. 6. The electron occupancy of $d_{xy}$, $d_{xz}$ and $d_{yz}$ orbitals of the orbitally equivalent V atoms (i.e. V1,V2,V5,V6) for solutions AFM(OS1) and AFM(OS2) as a function of $U$ within unconstrained LSDA+{\it U} calculations. Figures (a, b) and (c, d) corresponding to fully localized (FL) and around mean field (AMF) double counting (DC) schemes, respectively.

FIG. 7. The difference of the electronic occupancy of $d_{xz}$ and $d_{yz}$ orbitals ($\mid$$d_{xz}$-$d_{yz}$$\mid$) of the orbitally equivalent V atoms (i.e. V1,V2,V5,V6) for AFM(OS1) and AFM(OS2) solutions with varying $U$ for fully localized (FL) and around mean field (AMF) double counting (DC) schemes.

\begin{figure}
\caption{}
\includegraphics{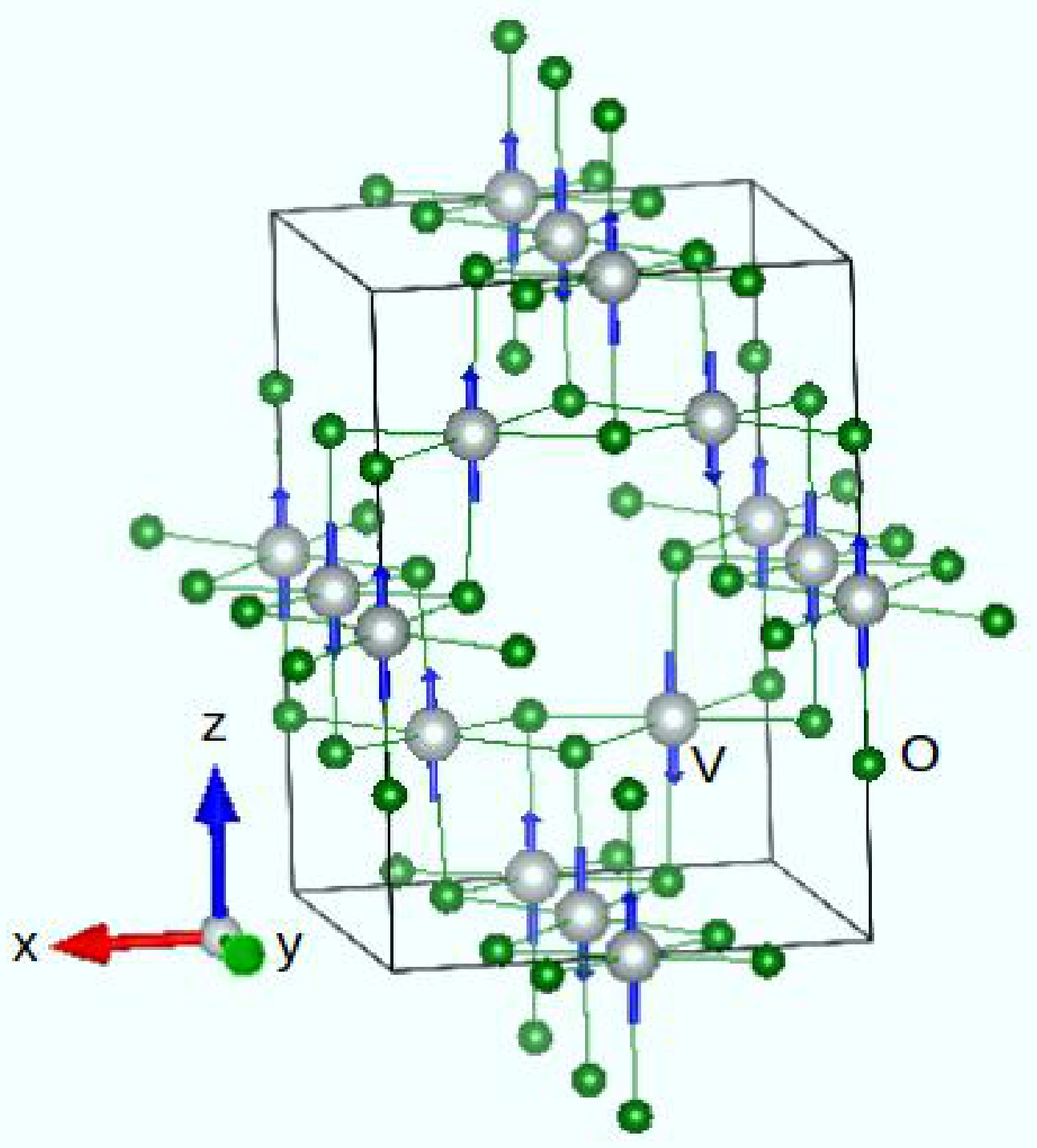}
\end{figure}

\begin{figure}
\caption{}
\includegraphics{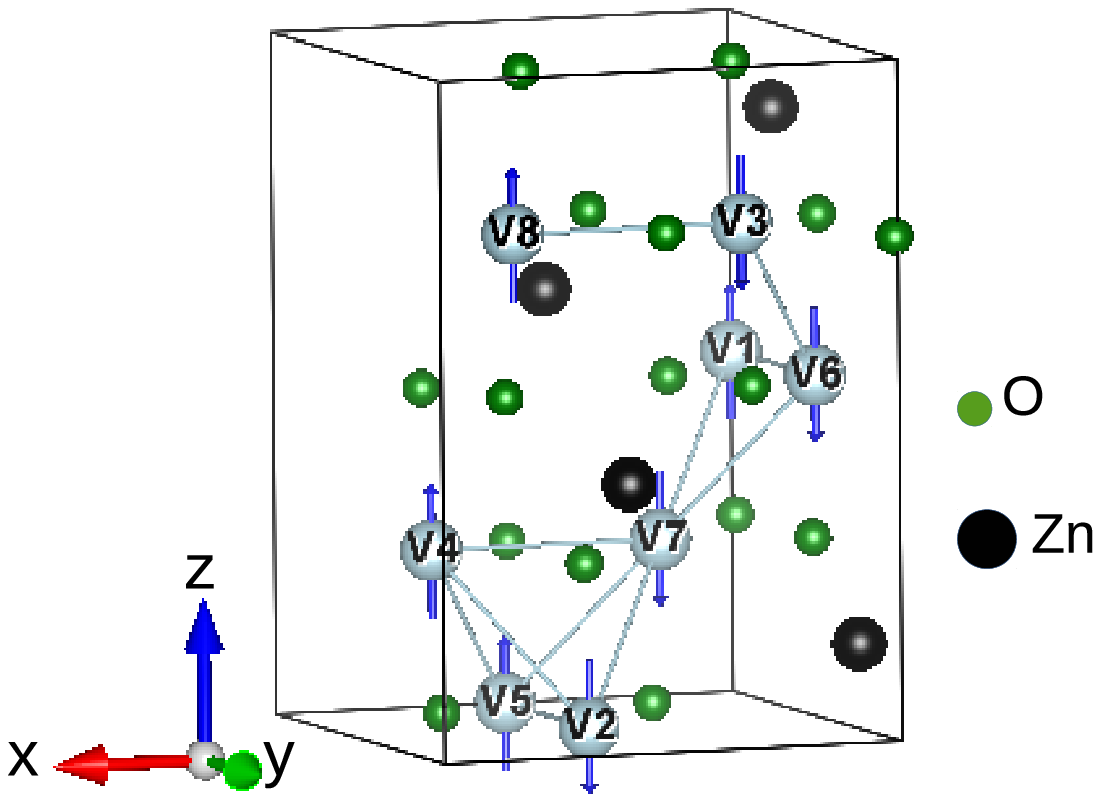}
\end{figure}

\begin{figure}
\caption{}
\includegraphics{Fig3.eps}
\end{figure}

\begin{figure}
\caption{}
\includegraphics{Fig4.eps}
\end{figure}

\begin{figure}
\caption{} 
\includegraphics{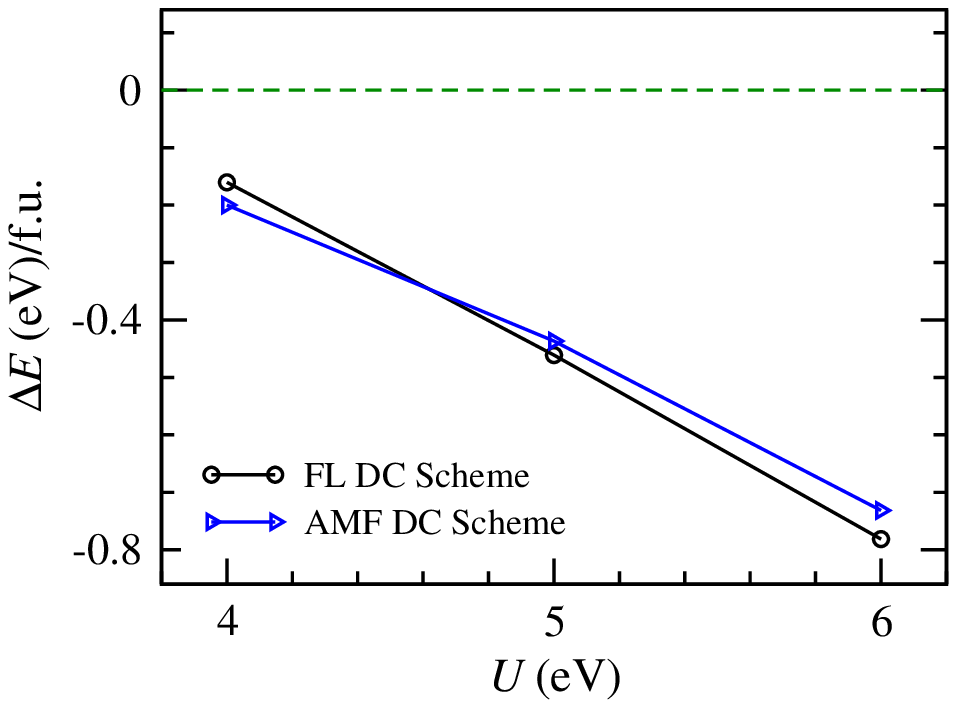}
\end{figure}

\begin{figure}
\caption{} 
\includegraphics{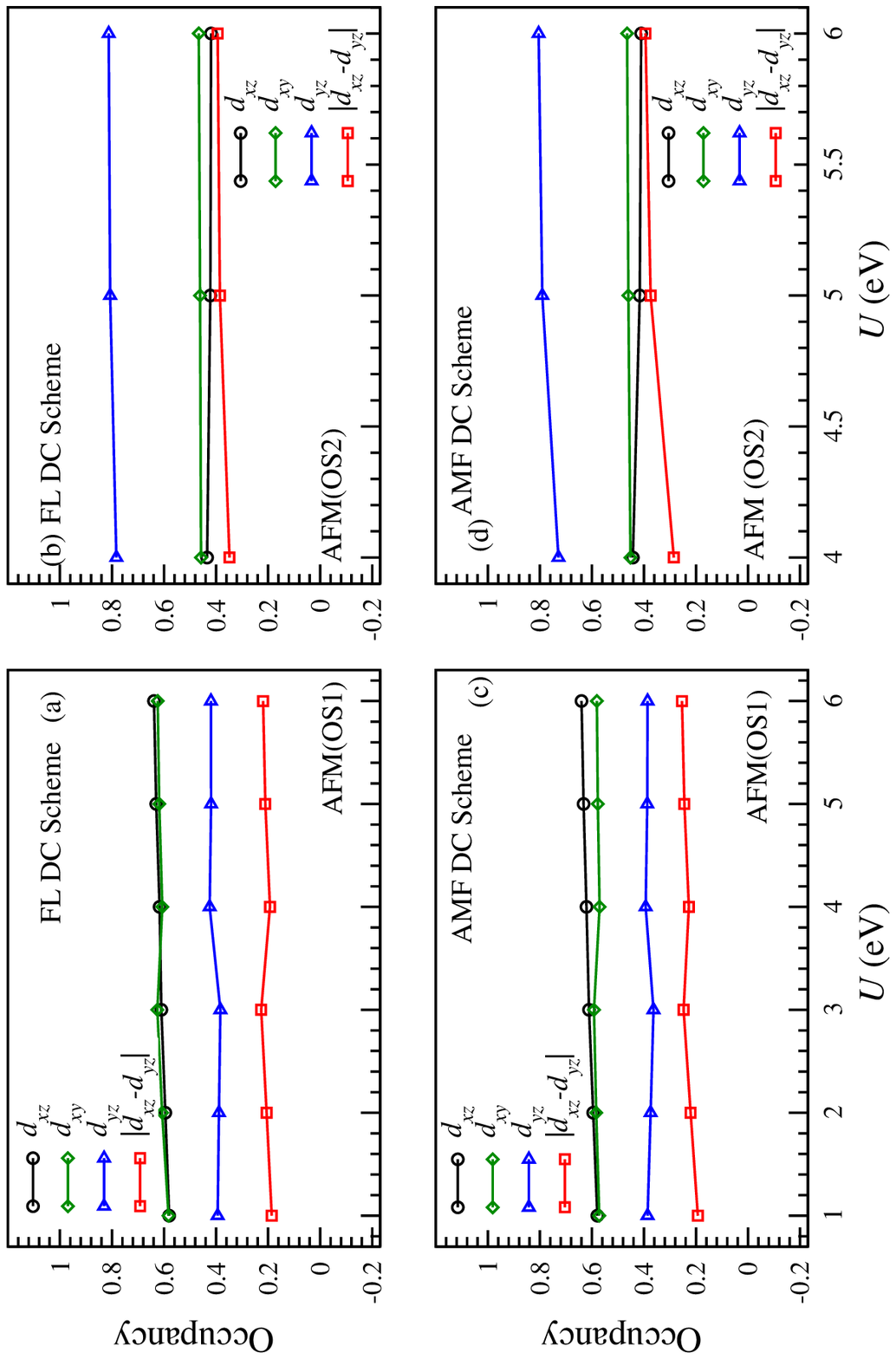}
\end{figure}

\begin{figure}
\caption{}
\includegraphics{Fig7.eps}
\end{figure}
       
\end{document}